\documentclass[%
 prd,
 twocolumn,
 nofootinbib,
 superscriptaddress,
 amsmath,amssymb,
 aps
]{revtex4-1}
\usepackage{amsmath, physics}
\usepackage{graphicx,xcolor}
\usepackage{dcolumn}
\usepackage{bm}
\usepackage{hyperref}

\begin{document}
$\left. \right.$
\begin{figure}[h]
\hfill\href{https://lisa.pages.in2p3.fr/consortium-userguide/wg_cosmo.html}{                                                     
\includegraphics[width = 0.17 \textwidth]{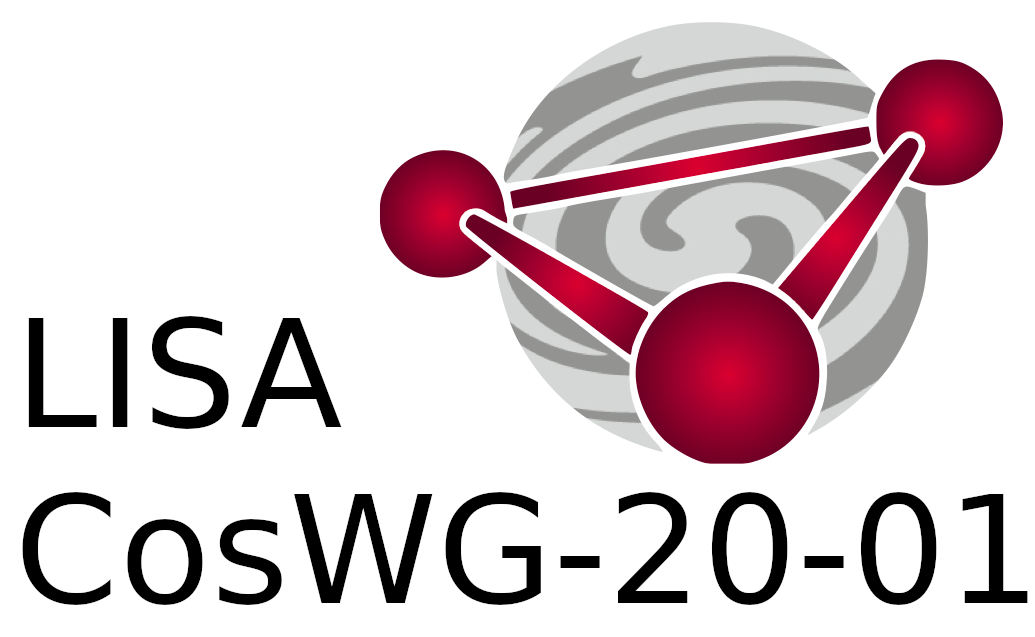}}
\end{figure}

\title{Maximum likelihood map-making with the Laser Interferometer Space Antenna}
\author{Carlo R. Contaldi}
\affiliation{Blackett Laboratory, Imperial College London, SW7 2AZ, UK}
\author{Mauro Pieroni}
\affiliation{Blackett Laboratory, Imperial College London, SW7 2AZ, UK}
\author{Arianna I. Renzini}
\email{Project coordinator and corresponding author\\ arianna.renzini15@imperial.ac.uk}
\affiliation{Blackett Laboratory, Imperial College London, SW7 2AZ, UK}
\author{Giulia Cusin}
\affiliation{Astrophysics Department, University of Oxford, DWB, Keble Road, Oxford OX1 3RH, UK}
\affiliation{Université de Genève, Dèpartement de Physique Thèorique and Centre for Astroparticle Physics, 
24 quai Ernest-Ansermet, CH-1211 Genève 4, Switzerland}
\author{Nikos Karnesis}
\affiliation{Laboratoire Astroparticules et Cosmologie, Université Paris Diderot,
10 Rue Alice Domon et Léonie Duquet, 75013 Paris, France}
\author{Marco Peloso}
\affiliation{INFN, Sezione di Padova, 35131 Padova, Italy}
\author{Angelo Ricciardone}
\affiliation{INFN, Sezione di Padova, 35131 Padova, Italy}
\author{Gianmassimo Tasinato}
\affiliation{Department of Physics, Swansea University, Swansea, SA2 8PP, UK}

\collaboration{For the LISA Cosmology Working Group}

 

\date{\today}


\begin{abstract}
Given the recent advances in gravitational-wave detection technologies, the detection and characterisation of gravitational-wave backgrounds (GWBs) with the Laser Interferometer Space Antenna (LISA) is a real possibility.
To assess the abilities of the LISA satellite network to reconstruct anisotropies of different angular scales and in different directions on the sky, we develop a map-maker based on an optimal quadratic estimator. The resulting maps are maximum likelihood representations of the GWB intensity on the sky integrated over a broad range of frequencies. We test the algorithm by reconstructing known input maps with different input distributions and over different frequency ranges. We find that, in an optimal scenario of well understood noise and high frequency, high SNR signals, the maximum scales LISA may probe are $\ell_{\rm max} \lesssim 15$. The map-maker also allows to test the directional dependence of LISA noise, providing insight on the directional sky sensitivity we may expect.      
\end{abstract}

\pacs{Valid PACS appear here}

\maketitle

\section{Introduction}
The Laser Interferometer Space Antenna (LISA) is an ESA led mission in collaboration with NASA planned to launch in the mid 2030s which will allow us to tune into a new, vast range of the gravitational-wave (GW) sky, at unprecedented depth and volume. \\
The LISA sensitivity curve spans roughly 4 orders of magnitude in frequency, from $10^{-5}$ Hz to $10^{-1}$ Hz \cite{Amaro2017}, with maximum sensitivity around $10^{-3}$ Hz, and is such that we may observe loud GW events out to redshift $z = 6$ and beyond \cite{Tamanini2016}.
In this range, LISA is expected to measure a motley of signals with respective signal-to-noise ratios (SNRs) ranging from a few hundred to a few, such that data analysis will require careful component separation to disentangle the individual coherent sources. There will also be the need for tools to analyse the stochastic signals which build up incoherently in the time stream.\\
The collection of incoherent, unresolved signals in the data are typically referred to as the gravitational-wave background (GWB), which is considered stochastic in the limit where there is a statistically significant set of uncorrelated overlapping waves.
There are several GW sources which will contribute to a stochastic background in the LISA band. These include galactic, extra-galactic or even primordial compact binaries, too distant or faint to be resolved \citep{Regimbau2011}, or a relic background from a burst of inflation or a phase transition at early times \citep{Grischuk1975,Maggiore1999,Hogan1986,Battye1997,Vilenkin2000, Caprini2018}. All backgrounds mentioned here are the focus of science objectives of the LISA mission~\cite{Amaro2017}. The detection of a cosmological GWB in particular is considered to be the ultimate challenge in cosmology as GWs are the only signal expected to reach us from before recombination and the consequent generation of the cosmological microwave background (CMB), providing unique insight on the origins of the Universe. 
In this {\it paper} we present a method to extract the stochastic signal in the LISA data and how to solve for its directionality, effectively mapping it on the sky. This will allow us to assess the angular resolution of the LISA detector for a distribution of extended, unresolved sources, as a function of SNR.

The mapper we have developed is based on an optimal quadratic estimator which solves for maps iteratively in the pixel domain. Our approach is influenced by work done previously with LIGO data, both as a part of the LIGO Virgo Collaboration~\cite{Abadie2011,LIGO2016a,TheLIGOScientificCollaboration2019} and as independent efforts~\cite{Renzini2019a,Renzini2019b}. Much work has been done to assess the ability of LISA to detect and characterise GWBs~\cite{Cornish2001, Cornish2017, Karnesis2019, Caprini2019, Smith2019}, and also study the sky response of LISA for map-making purposes~\cite{Kudoh:2004he,Romano2017}. Some studies have focused on how to distinguish detector noise from the galactic stochastic background, often referred to as a foreground when compared to other signals, which is expected to imprint a seasonal modulation on the time stream~\cite{adamscornish1,adamscornish2}. These are particularly relevant here as we will probe the ability of LISA to map the galactic foreground in Section~\ref{sec:maps} by analysing an idealised signal with a Milky-Way--like sky distribution. Map-making with space-based detectors is discussed extensively in~\cite{Taruya:2005yf,Taruya:2006kqa}, both in the low frequency limit and at high frequencies where a higher resolution may be achieved. However, these approaches reduce the mapping problem to a least-squares solution and solve for the spherical harmonic coefficients of the signal by looking at the individual cross-correlations between data-streams. Additionally, there is work in the literature which proposes solving for phase-coherent backgrounds, which we find is not possible, due to both the characteristics of the signal and the nature of the measurement. In this work, we present the first all-sky analysis based on the full likelihood of the data given a GWB intensity signal.  The aim is to deliver the framework of the LISA SGWB map-making pipeline which will be progressively refined and updated until the data becomes available.

To test our procedure we will use injected signals from different input maps in order to verify the reconstruction of a known sky, with simple spectral dependence. The inputs will consist in stationary maps of strain intensity of varying amplitudes and with anisotropies given by a Gaussian random field. These are not to be regarded as realistic simulations of any particular GWB and are used solely for testing purposes. 

This \emph{paper} is organised as follows; in Section~\ref{sec:GWB} we review the strain signal of GWBs and detail the LISA detector response to the intensity of the GWB on the sky. In Section~\ref{sec:maps} we introduce the quadratic estimator used to obtain maps of the GWB with LISA, and present results of a number of mapping tests we have run. Finally, we conclude in Section~\ref{sec:concl} by giving an overview of the impact of our results and anticipate useful extensions of our algorithm to be explored in the future.

\begin{figure*}[t]
    \centering
    \includegraphics[width = 0.4 \textwidth]{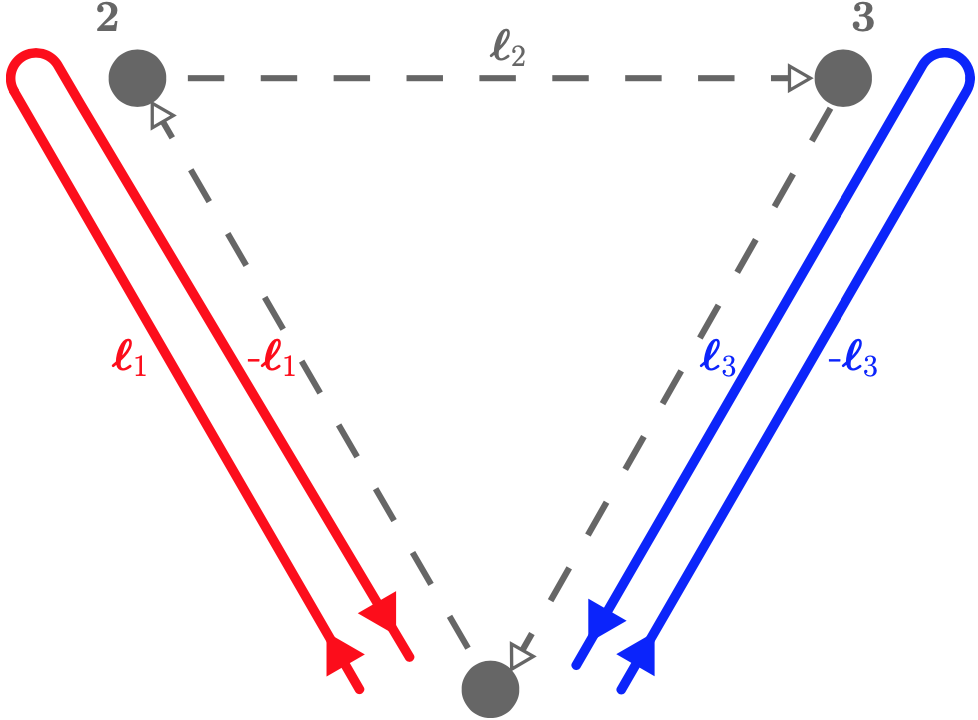}
    \quad\qquad\quad\qquad
    \includegraphics[width = 0.4 \textwidth]{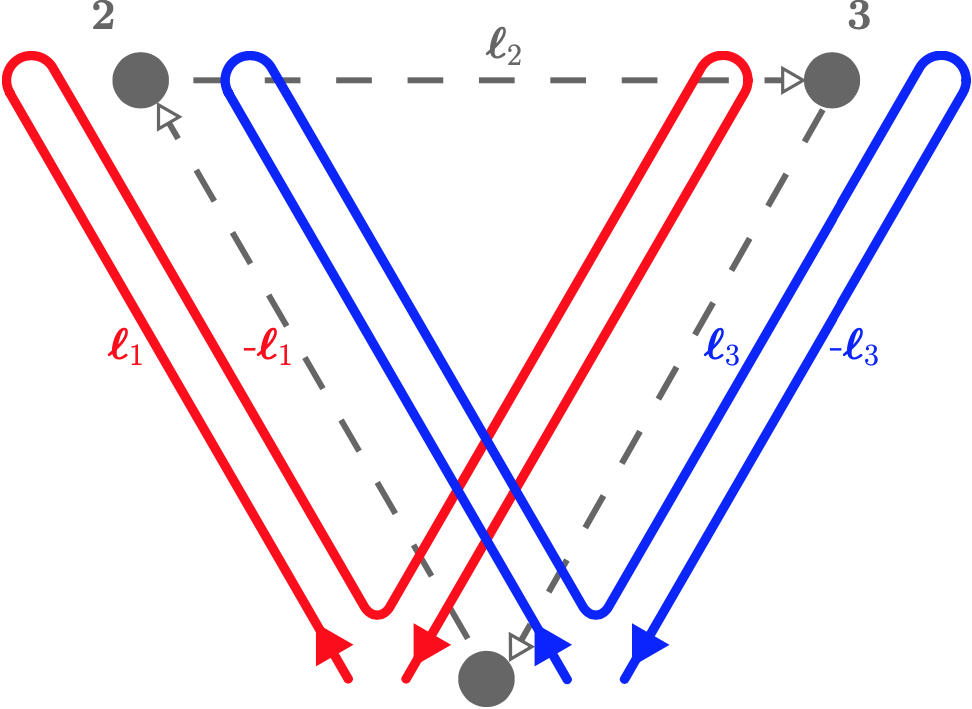}
    \caption{Illustrations of the TDI 1 (left panel) and TDI 1.5 configuration $X$ (right panel). The three numbered grey circles represent the spacecrafts arranged in an equilateral triangle. The links are labelled $\bm\ell_i$ in the left panel for clarity, and are the same as the ones in the right panel.}
    \label{fig:TDI}
\end{figure*}

\section{The signal of Gravitational Wave Backgrounds}\label{sec:GWB}
Typically the contribution of gravitational waves to the overall energy content of the Universe is parametrised by the dimensionless fractional energy density $\Omega_{\rm GW}$ \citep[see e.g.][]{Allen1999}. The spectral dependence of this measure is given by the physical energy density of GWs per logarithmic frequency interval,
\begin{equation}
\Omega_{\rm GW}(f) = \frac{1}{\rho_c} \frac{d\rho_{\rm GW}}{d\ln f}\,,
\end{equation}
where $\rho_c$ is the critical energy density. 
The fraction of this quantity produced by resolved GW sources is strictly marginal, since most of the signal is due to unresolved sources below the confusion limit. The collection of all unresolved GWs defines the GWB, which includes signals of both astrophysical and cosmological nature.

Whilst GWBs of different origin will have differing spectral dependence~\cite{Caprini2019}, most will be unpolarised and incoherent in that the temporal phase of the signal is expected to be random. The randomisation of the temporal phase will either be a product of the generation mechanism of the background, a consequence of propagation through the inhomogenous universe, or both~\cite{Cusin:2017mjm, Conneely:2018wis, Bartolo:2018evs, Bartolo:2019yeu, Bartolo:2019oiq, Margalit:2020sxp}.
Beyond the isotropic value of the GWB over the entire sky we can include any anisotropy by adding a directional dependence $\Omega_{\rm GW}(f) \rightarrow \Omega_{\rm GW}(f,\hat{\bm n})$ where $\hat{\bm n}$ is the unit vector of a line-of-sight on the sky.
Throughout this {\it paper} we will assume the frequency and sky dependence of the GWB are independent to good approximation, i.e.
\begin{equation}
    \Omega_{\rm GW}(f,\hat{\bm n}) \approx H(f)\,\tilde{ \Omega}_{\rm GW}(\hat{\bm n})\,;
    \label{OmegaEf}
\end{equation}
this applies both to the model of the incoming signal and the recipe for the signal's reconstruction. This is widely accepted to be a reasonable working assumption by the community, starting from \cite{Allen1996}. Indeed,  several models for GWBs predict this property to apply for the ensemble average of a stochastic GW signal. Specifically, the signal needs to be stochastic in both the time- and frequency- domain. It is worth noting that we may not be operating in the truly stochastic limit, however we will make this assumption throughout. In this paper, we further assume that the spectral shape of $\Omega_{\rm GW}$ is described by a power law with parameters $(\alpha, f_0)$ such that 
\begin{equation}
    \Omega_{\rm GW}(f,\hat{\bm n}) = \left(\frac{f}{f_0}\right)^\alpha\, \tilde{\Omega}_{\rm GW}(f_0,\hat{\bm n})\,,
    \label{eq:Omegazero}
\end{equation}
in allignment with both theory- and data- driven analyses of the stochastic GW signal~\cite{Christensen2018, Renzini2018, Smith2019}. Specifically for the monopole component of the stochastic background this assumption has been relaxed and a fitting technique has been tested using broken-power-law templates in the ground-based detector frequency band~\cite{Kuroyanagi:2018csn}. This fitting technique is very effective when tested on idealized signals, however the application to realistic data requires improvement as component separation will need to be taken into account. In the LISA case, a method to go beyond the single power-law assumption for the reconstruction of the monopole component $\Omega_{\rm GW}(f)$ has been proposed in~\cite{Caprini2019}.
We plan to explore the validity of~(\ref{eq:Omegazero}) in the future, by analysing a realistic, time-domain generated GWB signal from the LISA Data Challenge (LDC).
 

\subsection{Strain signal}\label{subsec:strain}

We model the GW strain just outside the detector with the transverse, traceless tensor $h_{ij}(t,\, {\bm x})$ at time $t$ and position $\bm x$. $h_{ij}$ can be decomposed into the independent polarisation states $h_+$ and $\,h_{\times}$ and expanded using plane waves as
\begin{equation}
h_{ij}\,(t,\bm{x})=\int_{-\infty}^{+\infty} \!\!\!df \int_{S^2} \!\!\!d\hat{\bm n}\!\!\sum_{P=+,\,\times}\!\!h_P\,(f,\,\hat{\bm n})\,e_{ij}^P(\hat{\bm n})\,e^{i2\pi f(\hat{\bm n}\cdot \bm{x}-t) }\,,
\label{expan}
\end{equation}
where polarisation base tensors $e^P$ are
\begin{align}
e^+ =e_\theta\otimes e_\theta - e_\phi\otimes e_\phi \,,&&\\
e^\times = e_\theta\otimes e_\phi+e_\phi\otimes e_\theta\,,&&
\end{align}
with
\begin{equation}
\begin{split}
e_\theta = (\cos\theta\cos\phi,\cos\theta\sin\phi,-\sin\theta) \,,\\    
e_\phi = (-\sin\phi,\cos\phi,0)\,,  
\end{split}
\end{equation}
as in~\cite{Renzini2018} and we are working in units where $c=1$.
For a stochastic GWB each Fourier mode will contribute independently to the overall signal, and $h_+$ and $h_\times$ are two random complex fields on the sky. 
We will assume the amplitudes are drawn from a Gaussian probability distribution. In the case of an astrophysical GWB this assumption holds
if the signal is sourced by a sufficiently large number of independent events and all high signal-to-noise outliers have been subtracted from the detector time streams, such that the central limit theorem applies~\cite{Ginat:2019aed}.
Under the Gaussian assumption, the statistical properties of the amplitudes are then characterised solely by the second order moments
$\langle h_P^{}(f,\,\hat{\bm n}) h^\star_{P'}(f',\,\hat{\bm n}')\rangle$, which, assuming statistical homogeneity, correspond to ensemble averages
\begin{equation}
\begin{split}
 & \begin{pmatrix}  \langle h^{}_+\,h'^\star_+\rangle  & \langle h^{}_+\,h'^\star_\times\rangle  \\ \langle h^{}_\times\,h'^\star_+\rangle  & \langle h^{}_\times\,h'^\star_\times\rangle  \end{pmatrix} = \frac{1}{2}\,\delta(\bm{n-n'})\,\delta(f-f')\,\times\\
 &\begin{pmatrix} I(f,\,\hat{\bm n})+Q(f,\,\hat{\bm n}) & U(f,\,\hat{\bm n})-iV(f,\,\hat{\bm n}) \\ U(f,\,\hat{\bm n})+iV(f,\,\hat{\bm n}) & I(f,\,\hat{\bm n})-Q(f,\,\hat{\bm n}) \end{pmatrix}\,,
\end{split}
\label{stokey}
\end{equation}
where we have introduced the Stokes parameters $I$, the intensity, $Q$ and $U$, giving the linear polarisation, and $V$, the circular polarisation. The four Stokes parameters completely describe the polarisation of the observed signal in analogy with electromagnetic Stokes parameters for the photon. The difference here is that whilst the electromagnetic $Q$ and $U$ Stokes parameters transform as spin-2 quantities with respect to rotations, their strain counterparts transform as spin-4 under rotations. In both cases the intensity $I$ behaves as a scalar under rotations.


\begin{figure*}[t]
    \centering
    \includegraphics[width = 0.31 \textwidth]{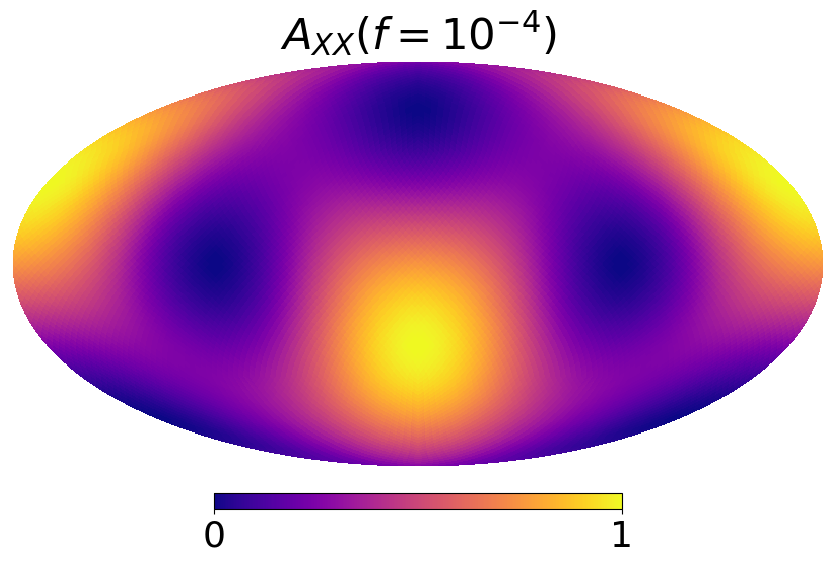}
    \hfill
    \includegraphics[width = 0.31 \textwidth]{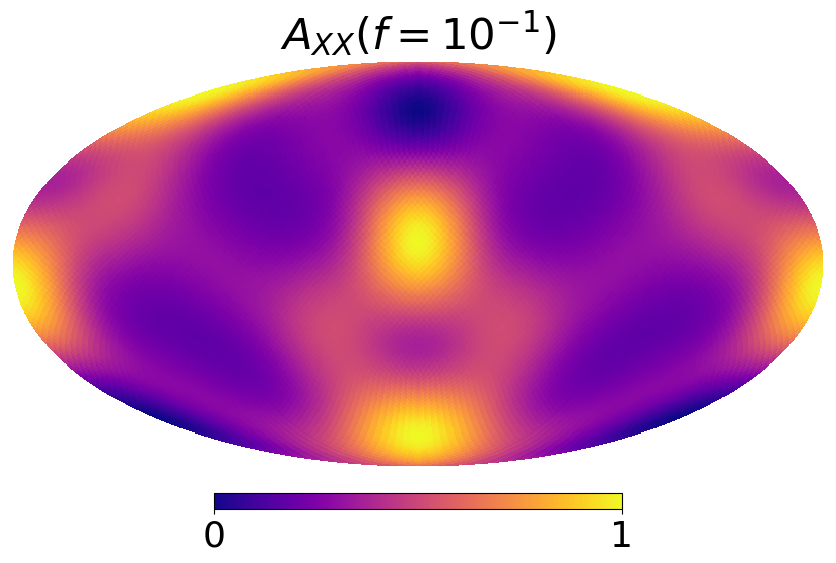}
    \hfill
    \includegraphics[width = 0.31 \textwidth]{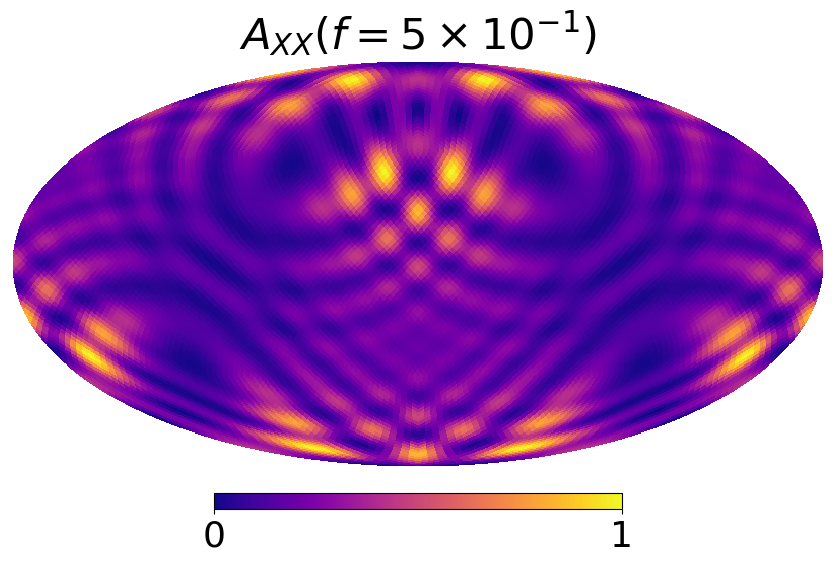}
    \caption{Normalised auto-correlated response of TDI channel X, $\bm A_{XX}$, at time $t = 0$ and in the Solar System Baycentre (SSB) reference frame, at frequencies $f = 10^{-4}$ Hz, $f = 10^{-1}$ Hz, $f = 5\times10^{-1}$ Hz from left to right respectively.}
    \label{fig:response}
\end{figure*}

In the following we will restrict our analysis to the reconstruction of the GWB intensity $I(f,\,\hat{\bm n})$ only, corresponding to the combination
\begin{equation}
I(f,\,\hat{\bm n}) = \, \langle h^{}_+(f,\,\hat{\bm n})\,h^\star_{+}(f,\,\hat{\bm n})\rangle  + \langle h^{}_\times(f,\,\hat{\bm n})\,h^\star_{\times}(f,\,\hat{\bm n})\rangle \,,
\label{Iishphc}
\end{equation}
note that this relates to the normalised logarithmic energy density $\Omega_{\rm GW}(f)$ as~\cite{Allen1996}
\begin{equation}   
\Omega_{\rm GW}(f) = \frac{4\pi^2f^3}{G \rho_c}I(f)\,,
\label{omegatoI}
\end{equation}
such that the assumption in Eq.~\ref{eq:Omegazero} carries down to the intensity
\begin{equation}
    I(f,\hat{\bm n}) = \left(\frac{f}{f_0}\right)^{\alpha-3}\, I(f_0,\hat{\bm n})\,.
    \label{Izero}
\end{equation}
This approximation is valid and sufficient for our calculations here, and has been assumed in multiple analyses of the stochastic background. Typically a value of $\alpha = 2/3$ is assumed for inspiral-dominated backgrounds~\cite{Sesana2008}, whereas $\alpha = 0$ may be used to describe scale-invariant cosmological backgrounds~\cite{Caprini2018,Contaldi:2016koz}.

\subsection{Detector Response}\label{ssec:DecRe}
The antenna is comprised of three spacecrafts which will
be positioned on three heliocentric yearly orbits in such a way that they will maintain an equilateral triangle
configuration throughout the whole duration of the mission (4+ years) \cite{Amaro2017}. The spacecrafts will specify a detector plane
at a $60^\circ$ angle with the ecliptic consistently throughout the motion of the constellation, which will appear to rotate on
the detector plane with period 1 year. Each pair of spacecrafts will share two laser links, which we can represent as two opposite vectors along the virtual detector arm. The arm-length will be time dependent during the mission, however in this work, for the sake of simplicity, we assume equal and stationary arms. Consequently, we consider equal and opposite links along each arm. In this paper we will use the trajectories as defined in~\cite{LISAdoc}, which correspond to analytical elliptical orbits in the Solar System Barycentre (SSB) frame. The starting position at zero time is chosen such that $\bm\ell_2$ in Fig.~\ref{fig:TDI} is aligned with the negative $y$ axis of the frame, and the two remaining arms are determined accordingly.

GW detection with LISA will rely on time-delay interferometry (TDI), which involves time-shifting and linearly combining independent Doppler measurements~\cite{Tinto2004}. This will occur in post-processing and will require accurate modelling of the response of each link. The study of the optimal TDI combinations, or channels, for GW detection is ongoing \cite{Muratore2020}. Given 6 links it is possible to construct three correlated channels, referred to as ${X,\,Y,\,Z}$, or two independent channels, typically referred to as ${A,\,E}$ channels, plus a {\it Sagnac} channel $T$~\cite{Prince:2002hp}.
The independent channel basis is found simply by diagonalising the correlated basis and rotating into the diagonal TDI space. As the measurements are time-delayed, this rotation will never be perfect and it adds a degree of complication which may be hard to constrain, hence we find it is preferable to work in the fully correlated basis, with the caveats it entails.
Specifically, the rotation will only cancel out the correlations assuming the three arm-lengths are equal, and that the noise in each spacecraft is identical. While these assumptions are made in this work, the method developed here is general and may be easily extended to accommodate a more complete noise model and detector response.
The simple Michelson-Morley-style combination is referred to as TDI 1 and it features three channels, each of which is centered around a spacecraft such that $X$ is centered around spacecraft 1, $Y$ around spacecraft 2, and $Z$ around 3. A similar, slightly more sophisticated combination designed to minimise the spacecraft breathing noise is referred to as TDI 1.5; both are illustrated in Figure~\ref{fig:TDI}. Throughout this paper we will adopt the TDI 1.5 configuration.

The time stream $s_C(t,\bm{x_i})$ measured by a single 1.5 TDI channel $C$ at time $t$ and $i$th spacecraft position $\bm{x_i}$ can be Fourier expanded between $t$ and $t+\Delta t$ to yield the signal as a function of frequency,
\begin{equation}
    s^\tau_C(f) = \sum_P \int_{S^2} d\hat{\bm n} \, R^P_C(f,\hat{\bm n};\tau) \, h^P(f, \hat{\bm n})\,,
\end{equation}
where $\tau$ is the time segment label, $R^P_C$ is the polarisation response function for the TDI channel $C$ and $h^P(f, \hat{\bm n})$ is the incoming GW strain decomposed into its polarisation components. We will now drop the $\tau$ label for simplicity.  The response functions for TDIs 1.5, $\{X,Y,Z\}$ are derived from the \textit{LDC Manual} which is an internal document. In the case of TDI 1.5 $X$ the measured strain may be written as   
\begin{equation}
    s_{X}(f) = -2i \sin(a) \, e^{-i a} \left[ e^{-i a} \left( y_{\bm\ell_1} - y_{-\bm\ell_3} \right) + y_{-\bm\ell_1} - y_{\bm\ell_3} \right]\,,
    \label{strain}
\end{equation}
where the $y_{\hat{\bm \ell}_i}$ terms correspond to the strain measured along the oriented link $\bm \ell_i$ and the exponential terms correspond to the phase shifts in Fourier space required to build the TDI 1.5 channel. Here $a = 2\pi f L$ for the sake of conciseness.  
The response of each link is
\begin{equation}
y_{\bm\ell_i} =
 - \frac{ia}{2}\, \sum_P \int_{S^2} Q^P_{\hat{\bm \ell}} h^P \, e^{- \frac{ia}{L} \hat{\bm n}\cdot{\bm x_i}} \, \text{sinc}\left( b \right)e^{-i b} \,,
 \label{yequ}
 \end{equation}
where $b = \frac{a}{2} (1-\hat{\bm n}\cdot\hat{\bm\ell_i})$ and the response of a single arm $\bm\ell$ to $P$- polarised modes, $Q^P_{\hat{\bm \ell}}$, is simply the contraction of the arm tensor with the polarisation basis element,
\begin{equation}
Q^P_{\hat{  \bm \ell}} = \epsilon^P : \hat{\bm \ell} \otimes \hat{\bm\ell}\,.
\label{Qp}
\end{equation}
Plugging (\ref{yequ}) into (\ref{strain}) we recover the sky response
\begin{equation}
    R^P_X =  \frac{ia}{2} \left( 1-e^{-2ia} \right) e^{-\frac{ia}{L}  \hat{\bm n}\cdot\vec{x}_1} \left[  Q^P_{\hat{\bm \ell}_1} \mathcal{T}(+\hat{\bm \ell}_1) - Q^P_{\hat{\bm \ell}_3} \mathcal{T}(-\hat{\bm \ell}_3)  \right]
\end{equation}
where the transfer function $\mathcal{T}$ to each sky direction $\hat{\bm n}$ is
\begin{equation}
\begin{split}
    \mathcal{T}(+\hat{\bm \ell}) &= \text{sinc}\left(\frac{a}{2} (1-\hat{\bm n}\cdot\hat{\bm \ell})\right)e^{- \frac{ia}{2} (3+\hat{\bm \ell}\cdot\hat{\bm n})} +\\
    &+\text{sinc}\left(\frac{a}{2} (1+\hat{\bm n}\cdot\hat{\bm \ell})\right)e^{- \frac{ia}{2} (1+\hat{\bm \ell}\cdot\hat{\bm n})}\,,
\end{split}
\end{equation}
assuming a permanent equal arm configuration, such that ${\bm x_i} = {\bm x_{i-1}}+{\bm \ell}_{i-1}$ where $i$ obeys cyclic permutations.
One can then derive $R^P_Y,\, R^P_Z$ by permuting the arms.

\begin{figure}[t]
    \centering
    \includegraphics[width = 0.45 \textwidth]{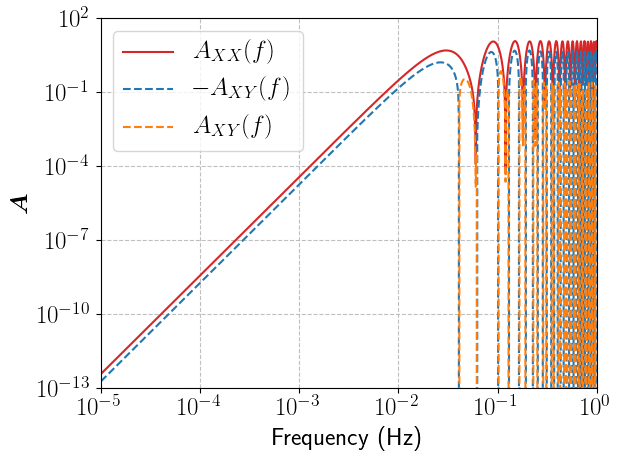}
    \caption{Sky-integrated auto-correlated and cross-correlated responses, $A_{XX}$ and $A_{XY}$, across the frequency spectrum. Note that the slopes in the low frequency limit scale as $\propto f^4$.}
    \label{fig:RespSpec}
\end{figure}

A signal composed of an incoherent superposition of many components, as in the case that we are focused on here, will vanish when averaged in time. To observe an incoherent GWB we therefore need to consider the integration of the square of the signal. Making use of all three channels we can write the TDI strain vector as $\bm s = (s_X, s_Y, s_Z)$ and construct the quadratic strain tensor $\bm S(f)$,
\begin{equation}
\bm S(f) = \bm s^{}(f) \otimes \bm s^\star(f)\,.
\end{equation}
We expect its ensemble average to yield
\begin{equation}
\langle \bm S(f) \rangle
= \int_{S^2} \,  d\hat{\bm n}\,\bm{A}(f,\,\hat{\bm n})\,I(f,\,\hat{\bm n})\,,
\label{corr}
\end{equation}
having expanded in the polarisation bases and inserting the relation between the second order moments of the strain and the Stokes parameters \cite{Allen1996}. Note that we consider a GWB which is stochastic also in polarisation; as such the $Q$, $U$, and $V$ Stokes parameters average to zero. The quadratic response tensor $\bm{A}$ is simply constructed with the linear response TDI vectors $\bm R^P = (R_X, R_Y, R_Z)^P$ 
\begin{equation}
    \bm{A}^\tau(f, \,\hat{\bm n}) = {\bm R}^+ \otimes {\bm R}^{+\star} + {\bm R}^\times \otimes {\bm R}^{\times\star}\,.
    \label{Atensor}
\end{equation}
We have recovered the $\tau$ label to point out the time-dependence of the detector response, which is implicitly expressed in Eqs.~(\ref{strain}-\ref{Atensor}) through the arm and positions vectors. Note also the directional and frequency dependence of $\bm{A}$, a sample of which is given in Fig. \ref{fig:response}.
Here the continuous sky response has been discretised to produce the Mollweide projection in pixel space, $\bm A (\hat{\bm n}) \to \bm A_{p}$, where $p$ labels a pixel on the sky, and it is shown in the SSB reference frame. The projections and pixel calculations are carried out using the HealPix package~\cite{Gorski:2004by}. As in the case of LIGO, the sky response gives rise to an inhomogeneous and non-compact antenna pattern. However, in the case of the LISA TDI channels this pattern is also significantly dependent on the frequency probed, as may be observed in the different panels of Fig. \ref{fig:response}. It is straightforward to show that in the low frequency limit the spectral dependence of the quadratic response tensor $\bm{A}$ scales as $f^4$ \cite{Caprini2019}, then starts deviating above $10^{-2}$ Hz, as may be observed in Fig.~\ref{fig:RespSpec}. The constant pattern and flat trend will set the limit to the resolution of LISA at low frequency, while this substantial deviation and the high-$\ell$ pattern will prove crucial to obtain higher resolution at high frequency.

%
%
    
For the simulations and study that follows we fix the coordinate system and polarisation basis to the SSB frame as it is convenient for the description of both the spacecraft trajectories and the injected signal's directional dependence. 
    

Although we have developed a map--maker that solves for a map directly in the pixel domain, we can use it to assess the ability of the cross-correlated TDI channels to reconstruct the angular power spectrum of a background defined as
\begin{equation}
    C_\ell^{\rm GW} = \frac{1}{2\ell + 1} \sum_{m = -\ell}^{\ell} \qty| \int_{S^2} \frac{d\hat{\bm n}}{4\pi}\, Y_{\ell m}(\hat{\bm n})\, I(\hat{\bm n}) |^2\,.
\end{equation}
It is then informative to decompose the detector response in spherical harmonic space, similarly to \cite{Renzini2018}, and assess the sensitivity to $C_\ell^{\rm GW}$ induced by the isotropised spherical harmonic response,
 \begin{equation}
     \bm A_\ell =  \frac{1}{2\ell + 1} \sum_{m = -\ell}^{\ell} \qty| \int_{S^2} \frac{d\hat{\bm n}}{4\pi}\, Y_{\ell m}(\hat{\bm n})\, \bm A(\hat{\bm n}) |^2\,.
 \end{equation}
Studying the relative sensitivity of a single TDI channel to angular scales $\ell$ on the sky, the angular resolution of the LISA TDI 1.5 configuration is predicted to be constrained as $\ell \lesssim 10$~\cite{Cornish2001,Smith2019}. This resolution will strongly depend on the frequency interval examined, such that for lower frequencies the cutoff will be much more severe. This discussion is expanded in~\cite{Angeloinprep}.  

\section{Maximum Likelihood Maps}\label{sec:maps}
\subsection{Mapping method}
In this section we describe the method we have developed to extract maximum likelihood maps of the GWB from three LISA TDI 1.5 data streams. We start with some remarks on analogies and differences with CMB analysis.

The final parts of the analysis of LISA data for the purpose of constraining GWBs will be split into two main stages. The first will be an estimate of a map in a way that is similar to the estimation of CMB maps from timestream data. This first compression step, from time (frequency) domain to sky coordinates, will be useful for identifying systematics in the data and for separating out different signals contributing to the GWB. 
The second stage will be a further compression to an isotropised statistic i.e. the angular power spectrum of the (separated) signal. This is similar to the power spectrum estimation step in CMB taking the maps to a final $C_\ell$ estimate.

In CMB observations made with incoherent (bolometric) detectors, the first stage uses a $\chi^2$ estimate to obtain a maximum likelihood map \cite{Borrill:1999kt}. This is possible because the signal component of the data does not vanish in the ensemble mean
\begin{equation}
    \langle d \rangle \equiv \langle s + n\rangle = m\,,
\end{equation}
where, generically, $d$, $s$, and $n$ are the time-domain data, signal, and noise with $\langle n\rangle$=0 and model $m\equiv \langle s\rangle$.
This allows the definition of a residual $d-m$ whose covariance is just the noise covariance $N\equiv \langle nn^\dagger\rangle$. This means we can estimate the signal by considering simply a Gaussian likelihood for the residual and, since the noise is usually considered to be independent of the signal, the normalisation of the likelihood is constant and we only need to minimize the $\chi^2$ part
\begin{equation}
    \chi^2(d|m) = (d-m)^\dagger N^{-1} (d-m)\,,
\end{equation}
to obtain a maximum likelihood solution for $m$ using the closed-form solution to $\partial \chi^2/\partial m=0$.

This is not possible for data obtained from noise correlated, coherent detectors such as LISA. In this case both signal and noise components in the time-domain vanish in the ensemble mean. Transforming to the Fourier domain does not alleviate this problem unless the background itself is phase-coherent or if the signal is a single point source on the sky. In this case no residual can be defined since $\langle d-m\rangle = 0$ and we need to use the full likelihood of the data with both $\chi^2$ and normalisation depending on the signal being estimated. There is no closed-form solution for the maximum likelihood in this case and iterative methods such as quadratic estimators must be used to estimate the signal \cite{Bond1998}. This is analogous to the {\sl second} stage in CMB analysis where the map signal component also vanishes in the ensemble mean limit and therefore the full data likelihood is used to estimate the signal. For LISA {\sl both} stages will require the use of a quadratic likelihood estimator and we define a method for the first stage below.

Note that a $\chi^2$ estimate is possible for LIGO style detectors where the noise can be assumed to be uncorrelated between detectors \cite{Renzini2019a, Renzini2019b, LIGO2016, TheLIGOScientificCollaboration2019}. In that case we can define a likelihood for a residual in the cross-correlated data where the signal component ensemble mean does not vanish but the noise does. In the case of LISA TDI $\{X,\,Y,\,Z \}$ channels, the noise will be correlated between channels and this approach is not possible.
 
Firstly, we define the likelihood for the data given a signal {\sl intensity}. We decompose the data TDI vector as $\bm{d}^{\tau}_{f} = \bm{R}h +\bm{n}$, where $\bm{n}$ is the noise TDI vector, $\bm{R}$ is the linear response TDI vector and there is an implicit sum over polarisations. Assuming the noise is zero–mean and Gaussian with covariance $\bm{N}_{f} = \bm{n} \otimes \bm{n}$ and the signal component is also Gaussian with covariance $\bm C^{\tau}_{f} = \bm A \tilde{I} + \bm N$, the likelihood $\mathcal{L}$ of a pixel map $\tilde{I}$ given the TDI data $\bm{d}$ is
\begin{equation}
    \mathcal{L} \propto \frac{1}{|\bm C|^{1/2}} e^{-\frac{1}{2}{\bm d} \, \bm C^{-1} \, {\bm d}^\star } \,, 
\label{Like}
\end{equation}
Following \cite{Bond1998} we find the solution which maximises $\mathcal{L}$,
\begin{align}
\tilde{I}_p &= \mathcal{F}_{pp'}^{-1} \cdot \text{Tr}\sum_{\tau, f} \left[ \bm C^{-1} \, \frac{\partial \bm C}{\partial I_{p'}} \, \bm C^{-1} \, (\bm D - \bm N) \right]\,, \label{eq:itersol}\\
 \mathcal{F}_{pp'} &= \text{Tr} \sum_{\tau, f} \left[ \bm C^{-1} \, \frac{\partial \bm C}{\partial I_{p}} \, \bm C^{-1} \, \frac{\partial \bm C}{\partial I_{p'}} \right]\,, \label{eq:fisher}
\end{align}

where we have explicitly written down the pixel indices for clarity. $\bm D = {\bm d} \otimes {\bm d}^\star$ is constructed thanks to the permuting property of the trace. $\mathcal{F}$ is the Fisher information matrix, whereas the trace in Equation~(\ref{eq:itersol}) is referred to as the gradient term.

This approach is general and may be used for both a broad- or a narrow- band analysis, and there is no specific limit on the time-discretisation as long as the Gaussian ansatz is not violated.
In the application considered here, the trace is taken over the three TDI channels, and the sum is over all frequencies in the FFT and all observation times, to maximise sky coverage. We can also consider integrating over short frequency intervals or single frequencies to probe the frequency dependence of the GWB, within the limits imposed by the conditioning of the Fisher matrix. We employ an iterative scheme to reach the maximum likelihood estimate for $I$, starting with an initial guess $\tilde{I}_{\rm in}$ to plug into $\bm C$, get a first estimate of $\tilde{I}_p$, then repeat until convergence. In the analysis here we assume to perfectly know the noise model $\bm N$, and expect that when this method will be applied to real LISA data it will be possible to rely on an independent noise estimation which informs the gradient term. However, it will be possible yet expensive to extend this estimator to include the noise as a free parameter, and solve for both noise and signal components directly.

In case  the signal is non-Gaussian, the maximum--likelihood maps derived below remain unchanged, but the interpretation of their covariance would be affected by the presence of higher order cumulants in the underlying probability densities. See e.g. \cite{Drasco:2002yd} for an  analysis of detection methods of non-Gaussianity in a GWB induced by short-duration signals, and \cite{Seto:2009ju} for a study of how to use higher-order cumulants to characterize its properties. 

It is important to stress that in this approach we assume that the intensity scales simply across the spectrum by some function $E(f)$. In reality this will not be the case and the model will have to account for a number of sky signals, each having a distinct spectral dependence. For a broad band analysis such as that carried out here one would have to solve for multiple maps. Alternatively, an analysis where the data is separated into narrow frequency bands, such as one of the approaches in \cite{Renzini2019b}, could deal with multiple signals through a later component separation stage.

\subsection{Mapping tests}

Here we present the results of the application of the method detailed above to simulated data of LISA TDI 1.5 $\{X,\,Y,\,Z\}$ channels. These serve as a proof of concept of the recipe, which is the first stepping stone towards testing it on mock LISA time streams.
We generate correlated data in the Fourier domain as $\bm D^\tau_f = \bm A I_{\rm in} + \bm N_{\rm in}$ over the period of one year divided in two-day segments for different frequency bands. This FFT time-scale determines the pixelisation resolution, as we consider the sky response to be stationary throughout the whole time segment, hence it is chosen to be sufficiently short to allow for a high $N_{\rm pix}$, but sufficiently long so as not to exclude the lower frequencies.
Each segment is heavily down-sampled in frequency to $\mathtt{N}_f = 200$ samples in order to lighten the computational load as much as possible without losing too much resolution in frequency, which would lead to a loss of information on the sky due to the coupling between $f$ and $\hat{\bm n}$ via $\bm A$. $\bm A I_{\rm in}$ is the signal component of the data contracted over pixels and $\bm N_{\rm in}$ is a Gaussian realisation of the noise model $\bm N$. 
The resulting data streams for the three channels are then plugged into Eq.~(\ref{eq:itersol}), and the Fisher matrix is iteratively computed, where the covariance matrix $\bm C$ is the full covariance of the data and simply $\frac{\partial \bm C}{\partial I_{p}} = \bm A_p$. All the terms in Eqs.~(\ref{eq:itersol} - \ref{eq:fisher}) are complex matrices and the reality condition imposed by the data is recovered when integrating over frequencies. 

The nominal SNR of the signal component of the data is estimated by integrating over observation time and frequency range as follows:
\begin{equation}
    \text{SNR} = \sqrt{\Delta T_{\rm obs} \int_{f_{\rm min}}^{f_{\rm max}} df\, \left(\frac{A I_{\rm in}}{N} \right)^2}\,,
    \label{eq:snr}
\end{equation}
where $A$ is the instantaneous response of channel $X$ and $N$ is the appropriate noise model.


\begin{figure}
    \centering
    \includegraphics[width = 0.45 \textwidth]{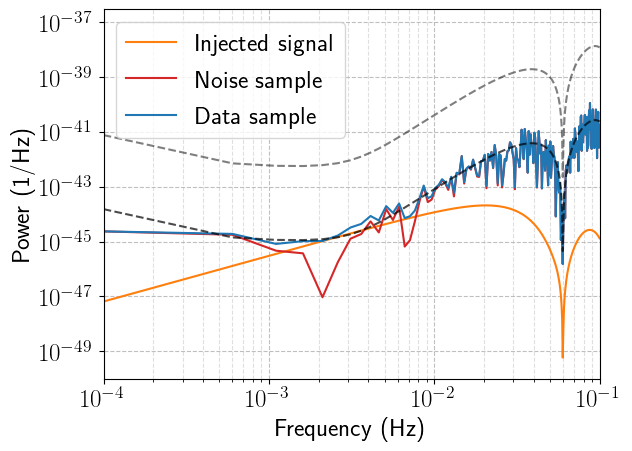}
    \caption{Data power sample from the autocorrelated X channel, decomposed into injected signal (orange line) and simulated noise (red line) components. The Injected signal has and SNR of 0.6, and is evidently buried in the noise. The sample represents the data accumulated over a two-day period and then FFTed. The dashed grey lines are the noise model which informs the noise generation; the darker line is renormalised by a factor which accounts for the lower sampling rate, while the lighter line is the true instantaneous noise curve used to calculate the SNR.} 
    \label{fig:datsam}
\end{figure}
\begin{figure*}[t]
    \centering
    \includegraphics[width = 0.31 \textwidth]{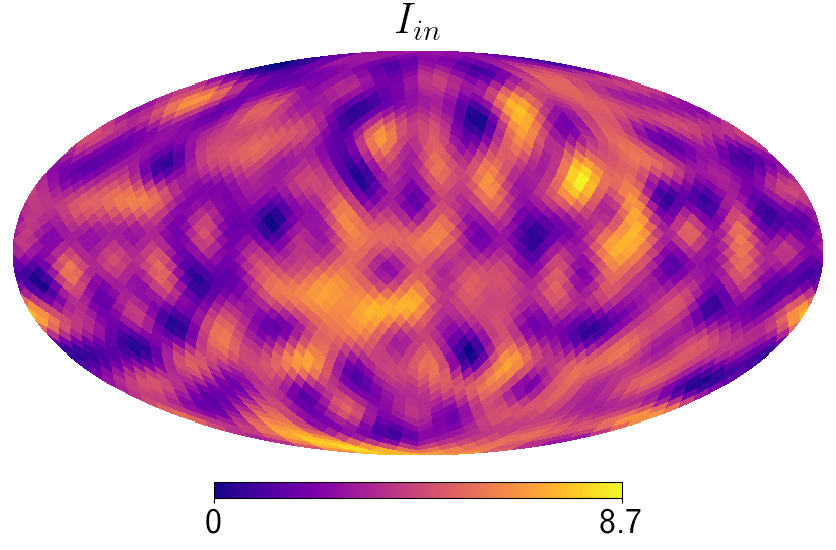}
    \hfill
    \includegraphics[width = 0.31 \textwidth]{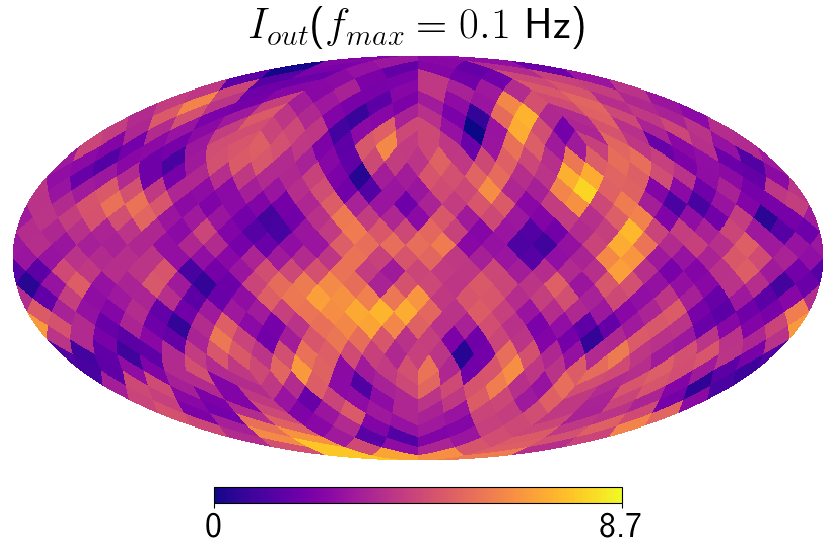}
    \hfill
    \includegraphics[width = 0.31 \textwidth]{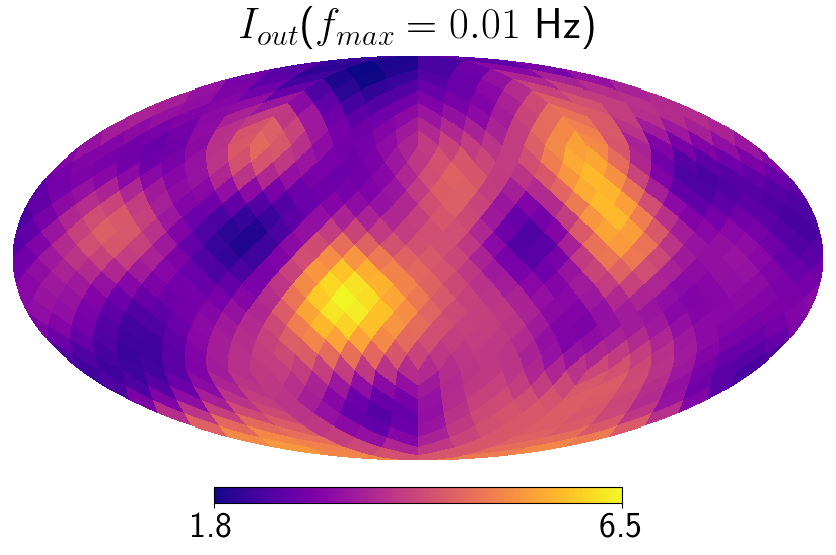}
    \caption{An example input map from the simulations (left panel) to be compared to the final output maps obtained integrating with different frequency cutoffs, $f_{\rm max} = 0.1$ Hz (central panel) and $f_{\rm max} = 0.01$ Hz (right panel). These highlight the different resolutions the LISA channels have in different ranges of frequency. }
    \label{fig:egmap}
\end{figure*}
\begin{figure*}[t]
    \centering
    \includegraphics[width = 0.48\textwidth]{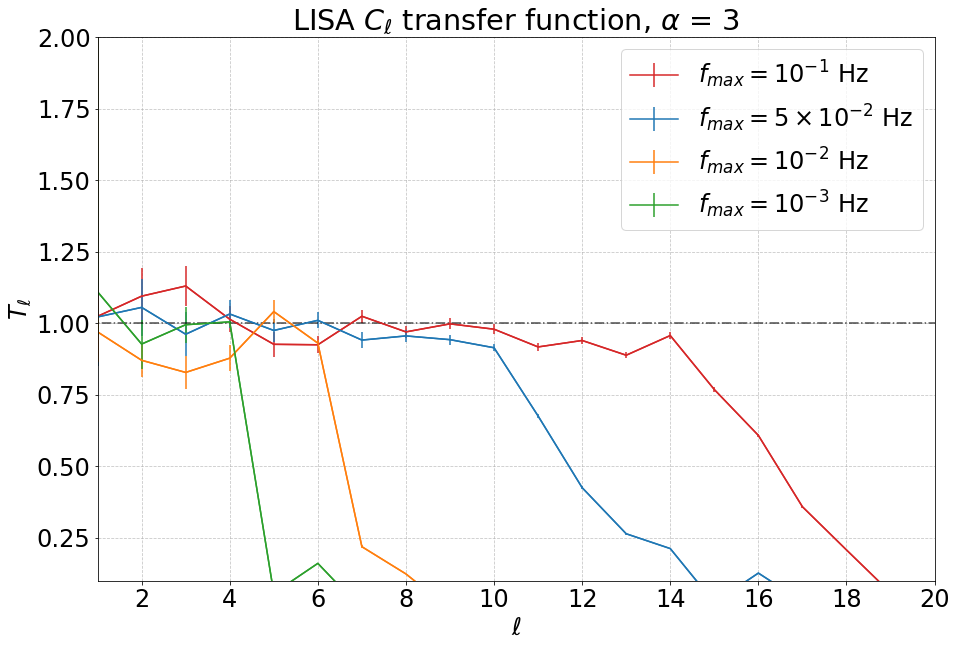}
    \hfill
    \includegraphics[width = 0.48\textwidth]{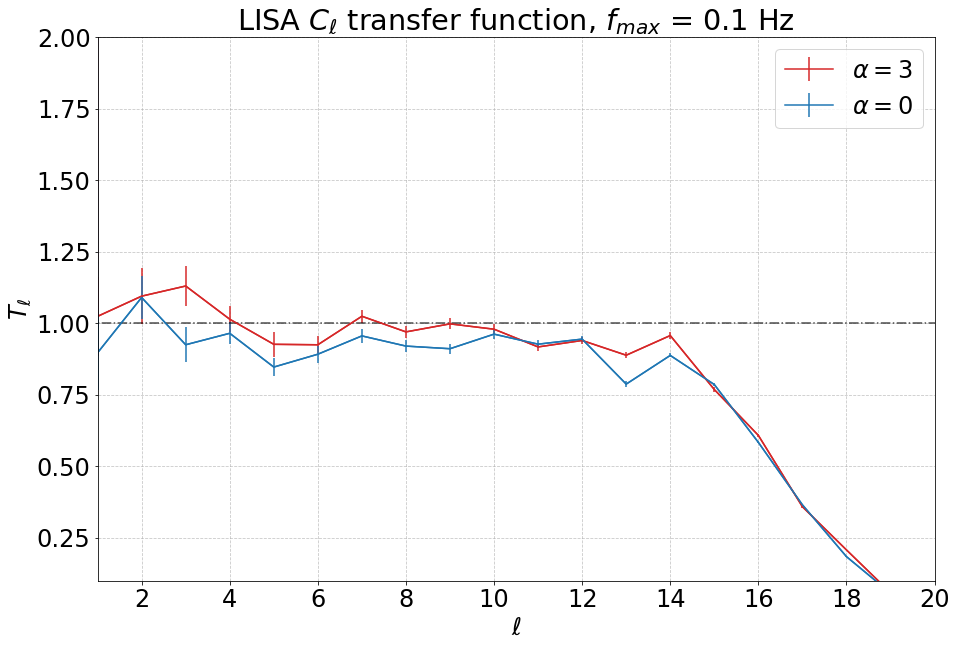}
    \caption{Transfer functions $T_{\ell}$ for the average reconstructed $C_{\ell}$s obtained with different frequency cutoffs (left panel) and different spectral shapes, $\alpha = 3$ and $\alpha = 0$, both in the high frequency case $f_{\rm max} = 0.1$ Hz (right panel). Each simulation set consists of 50 maps, each a different realisation of the same $C_{\ell}$ input. There appears to be a clear one-to-one relation between the resolution $\ell_{\rm max}$ of the instrument and the frequency cutoff. Conversely, there is an average difference of 5\% between the transfer functions obtained with different spectral weightings, however this does not affect the resolution cutoff.}
    \label{fig:fmax}
\end{figure*}
\subsubsection{Signal generation}
The signal component in the data $\bm A I_{\rm in}$ is generated by simply scanning an input map for the GWB intensity on the sky $I^p_{\rm in}(f_0)$ fixed at a reference frequency $f_0$ with the quadratic response of our TDI configuration directly in pixel space,
\begin{equation}
    \left(\bm A I_{\rm in} \right)^\tau_f = \frac{4\pi}{N_{\rm pix}} \sum_p \bm A^\tau_p (f) I^p_{\rm in} (f_0)\,,
\end{equation}
where we include the spectral dependence of the signal $E(f)$ in the response $\bm A$ for convenience.
This is a completely deterministic calculation which relies on the assumptions that the signal is truly stochastic and 
obeys Eq.~\ref{OmegaEf}, and that the overall integration over time and frequencies satisfies the ensemble average limit, such that Eq.~\ref{stokey} holds. All the maps presented here are reconstructed with number of pixels $N_{\rm pix}^{\rm out} = 768$ which corresponds to a HealPix $N_{\rm side} = 8$, while the injected maps are over-resolved with $N_{\rm pix}^{\rm in} = 3072$ which corresponds to a HealPix $N_{\rm side} = 16$.

Except where explicitly stated, the $I_{\rm in}$ maps scanned in the simulations below are random Gaussian realisations of an $\ell^2 C_\ell$ flat power spectrum with $\ell_{\rm max} = 20$, chosen uniquely for testing purposes.
This means that we are assuming that the monopole $C_0$ sets the size of the anisotropies which scale as a fixed power law $C_{\ell}=C_0/\ell.$  In a more realistic modeling of the signal, one should rather introduce two separate scales: a monopole $C_0$, which would set the size of the SNR as in~\ref{eq:snr}, and a typical scale of anisotropies $C_1$. These two scales are simply assumed to be the same here.
We explore the ability of LISA to reconstruct these input maps by changing the frequency integration range, varying the spectral parameter $\alpha$, and toggling the sky-integrated amplitude $I(f_0)$ with respect to the noise level set by $N$.
\begin{figure*}[t]
    \centering
    \includegraphics[width = 0.31 \textwidth]{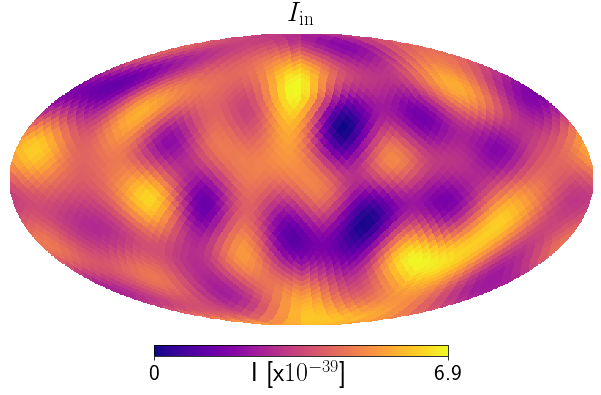}
    \hfill
    \includegraphics[width = 0.31 \textwidth]{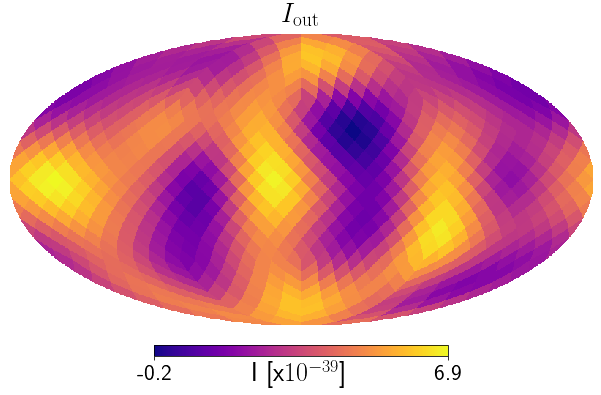}
    \hfill
    \includegraphics[width = 0.31 \textwidth]{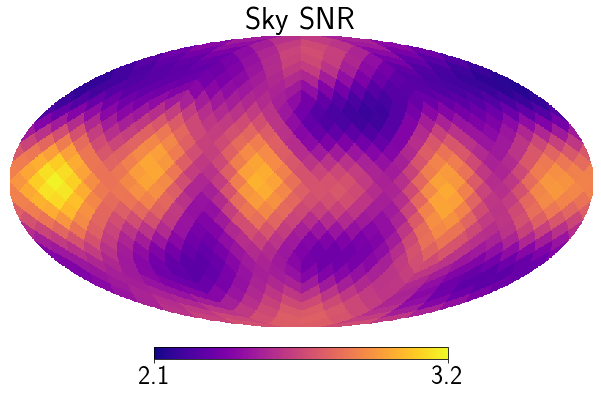}
    \caption{Input (left panel) and reconstructed output (central panel) maps in a low SNR case, integrating over a frequency range of $[10^{-4},\,10^{-1}]$ Hz. The input map is an $\ell C_{\ell}$ flat gaussian realisation with $\ell_{\rm max} = 10$. The output map is a heavily smoothed version of the input, with some loss of power due to the strong conditioning required. The SNR of this particular reconstruction is shown in the right panel.} 
    \label{fig:maps_w_noise}
\end{figure*}

\subsubsection{Noise generation}
LISA TDI noise is an active area of research, and there are ongoing studies on which TDI configuration will yield the lowest noise measurements. However these go beyond the scope of our project, as these model the orbits and breathing modes of the constellation more realistically.
In the simplified equal-arm scenario considered here, to good approximation the noise in each auto-correlation of TDI channels may be described by the same power spectrum, $S_{CC}$, and similarly the noise in each cross-correlation may be described by $S_{CD}$.
The $3\times3$ model TDI noise correlation matrix $\bm N$ is then completely described by $S_{CC}$ on the diagonal and $S_{CD}$ on the off-diagonal terms. The expressions for these power spectra used in this work are 
\begin{align}
    S_{CC} &= 16\sin^2{a} \, (S_{\rm int} + (3 +\cos{2a})S_{\rm acc})\,,\label{noisecurv1}\\
    S_{CD} &= -8\sin^2{a} \,\cos{a} \, (S_{\rm int} + 4 S_{\rm acc} )\,,
    \label{noisecurv2}
\end{align}
as presented in \cite{Smith2019}, where $S_{\rm int}$ and $S_{\rm acc}$ are the interferometer and acceleration noise-components respectively.

The noise realisation $\bm N_{\rm in}$ must respect the correlations imposed by $\bm N$, hence the noise is first generated linearly in the noise-diagonal space, and then rotated back into correlated noise space. This is achieved by generating a random 3-vector in noise-diagonal space and rotating it into the TDI noise vector $\bm n_{\rm in} = (n_X,\,n_Y,\,n_Z)_{\rm in}$ using the eigenvector matrix of $\bm N$. 
In the limit of equilateral configuration and identical noise at the vertices that we are considering here, the noise-diagonal space is precisely the space of the $\{A,\,E,\,T\}$ channels, and the same rotation as described in~\cite{Prince:2002hp} is employed to transform from one to the other.
 $\bm N_{\rm in}$ is then simply the outer product of $\bm n_{\rm in}$. An example data segment is provided in Fig.~\ref{fig:datsam}.

\subsubsection{Results}
We have run a variety of simulations to extensively probe the ability of the TDI configurations $\{X,\,Y,\,Z\}$ to reconstruct an anisotropic GWB, both in high and low signal-to-noise ratio (SNR) scenarios. For all the maps discussed below the inversion problem is strongly ill-conditioned, hence the Fisher matrix is inverted using the singular value decomposition technique as in similar work done with LIGO data~\cite{Renzini2019a, Renzini2019b}. Specifically, the condition number of the Fisher matrices in the cases presented below, which are all of dimension $N_{\rm pix}^{\rm out} \times N_{\rm pix}^{\rm out}$, is of order $10^{18}$. In the case of high SNR the output maps mildly depend on the conditioning imposed at pseudo-inversion as the signal will dominate the information, whereas in the low SNR scenario the output map is extremely dependent on the conditioning. In these tests, the choice made was to monitor the monopole level of the output map and gauge the conditioning such that it would match the input. This technique may be employed also with real data, where the measurement of the monopole can be done independently of the inversion problem.

To test the limits of the geometric set-up of LISA, we have generated data with anisotropic, flat spectrum backgrounds ($\alpha = 3$) of effectively infinite SNR and reconstructed output maps truncating the frequency integration at different values of $f_{\rm max}$. In this extremely high signal scenario, the solution converges after a single iteration. As may be seen in the example maps in Fig.~\ref{fig:egmap}, the input map is reconstructed remarkably differently in the case of $f_{\rm max} = 10^{-1}$ Hz and $f_{\rm max} = 10^{-2}$ Hz. The higher angular modes are well preserved when allowing the reconstructor to integrate up to higher frequencies, where there is finer structure in the response pattern, whereas they are aliased into lower modes when integrating only over the lower frequencies. 
To study this trend we have run sets of 50 analogous simulations when $f_{\rm max} = 10^{-3}$ Hz, $10^{-2}$ Hz, $5\times10^{-2}$ Hz, $10^{-1}$ Hz and calculated the average output $C_{\ell}$s in each set. We also compute the transfer functions $T_{\ell} = C_{\ell}/C^{\rm in}_{\ell}$. The comparison between
transfer functions with different frequency cutoffs may be seen in the left panel of Fig.~\ref{fig:fmax}.
\begin{figure}
    \centering
    \includegraphics[width = 0.45 \textwidth]{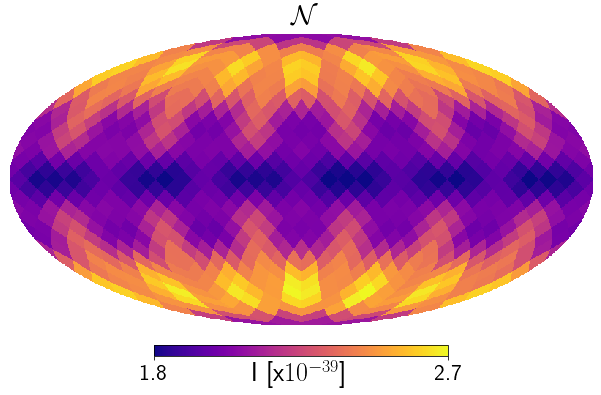}
    \caption{Sky distribution of the noise in the SSB frame. Note the imprinted 6-fold symmetry of the orbit, given by the three concentric orbits of the spacecrafts.} 
    \label{fig:noise_map}
\end{figure}
\begin{figure*}[t]
    \centering
    \includegraphics[width = 0.45 \textwidth]{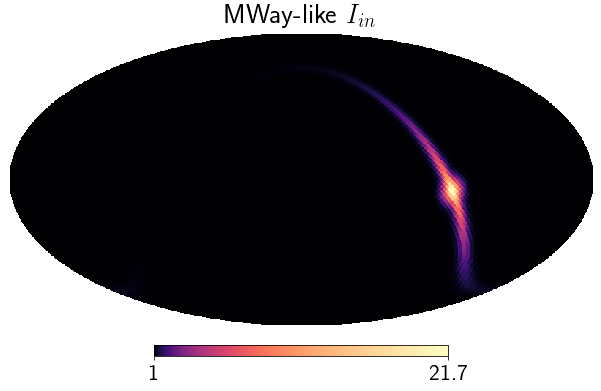}
    \hfill
    \includegraphics[width = 0.45 \textwidth]{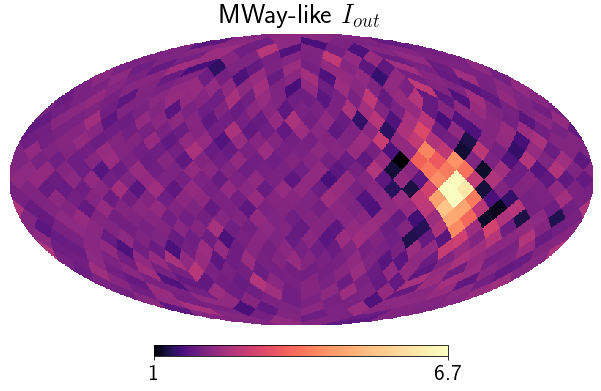}\\
    \includegraphics[width = 0.45 \textwidth]{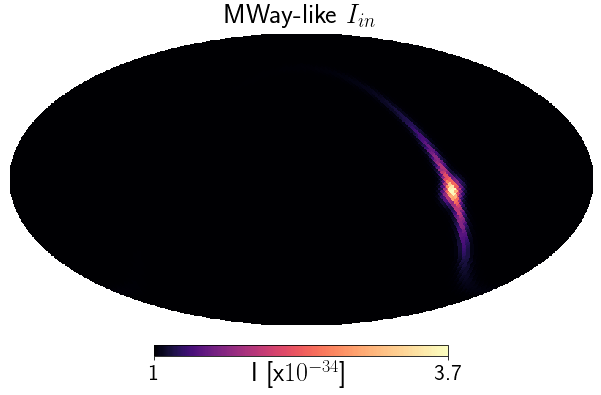}
    \hfill
    \includegraphics[width = 0.45 \textwidth]{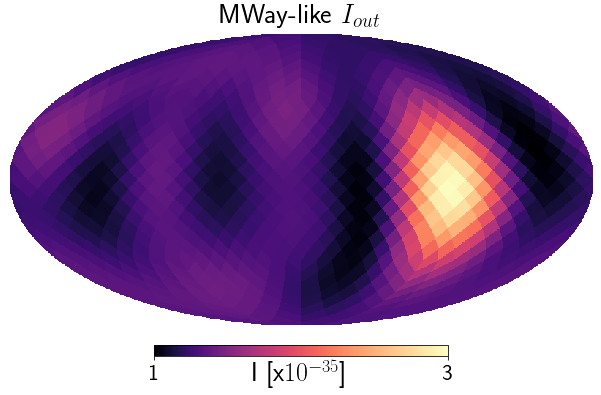}
    \caption{Sky inputs (left panels) and reconstructed outputs (right panels) for a Milky Way - like GWB distribution. The signal is injected in the range $[10^{-4}, 5\times10^{-3}]$ Hz. The top row is an extremely high SNR case, whereas the bottom row is an SNR 52, hence it emerges above the noise around $10^{-3}$ Hz. Both results are obtained over a full year of integration. These Mollweide projections are presented in log scale. }
    \label{fig:MWay_maps}
\end{figure*}
Different spectral shapes are reconstructed slightly differently, as may be seen in the right panel of Fig.~\ref{fig:fmax} where the high frequency $\alpha = 3$ set is compared with a high frequency $\alpha = 0$ set of simulations. The $C_{\ell}$ reconstruction and resulting transfer functions differ at certain modes, on average by 5\%, but the cutoff $\ell_{\rm max}$ appears to be the same, hence we conclude the frequency weighting in the signal and reconstruction do not have a strong impact on the resolution of the reconstructed maps.

An example of map reconstruction in the presence of high noise is shown in Fig.~\ref{fig:maps_w_noise}. The map in input here is a $\ell C_{\ell}$ flat Gaussian realisation with maximum angular scale $\ell_{\rm max} = 10$ to make the visual comparison easier. The spectral shape of the signal is $\alpha = 2/3$, which together with the power spectrum is typically associated with an astrophysical inspiral-dominated GWB~\cite{Sesana2008} which traces the large scale structure~\cite{Cusin:2018}.
The signal component has an SNR of 0.6 and is thus buried under the noise, as may be seen in the specific data sample in Fig.~\ref{fig:datsam}.
As may be observed in Fig.~\ref{fig:maps_w_noise}, the presence of loud noise has a similar effect as having a lower frequency cutoff, as the noise dominates at higher frequencies. Additionally, severe conditioning is required to cut out the noisier high modes, and multiple iterations are necessary to reach convergence. Specifically, the final conditioning cuts 97\% of the total eigenvalues of the Fisher information matrix. The value of the monopole of the reconstructed map however does not depend strongly on the conditioning and is in good agreement with that in input. Thus the resulting heavily conditioned output map appears to be a low-$\ell$ smoothed version of the initial map with some loss of power. We also recover the noise distribution on the sky $\mathcal{N}$, shown in Fig.~\ref{fig:noise_map}, as the diagonal of the inverse Hermitian square root of the converged Fisher matrix. This is an incomplete representation of the noise as the off-diagonal correlations are ignored, however it already shows that there is more noise off the ecliptic, as expected.

Finally, an example of reconstruction of a galaxy-like sky signal both with and without noise over a full year of integration is shown in Fig.~\ref{fig:MWay_maps}. The input signal has spectral shape $\alpha = 2/3$ in the frequency range $[10^{-4},5\times10^{-3}]$ Hz, as suggested in~\cite{Nelemans2001}, and the directional dependence is that of the Milky Way's white dwarf binary (WDB) population from~\cite{Korol2017}.  
The high SNR case is extremely unrealistic but provides a useful visual test of what the best resolution of LISA can be for a confused GWB at the frequency range where the signal from the compact galactic WDB is dominant. By relaxing the conditioning and including the higher modes in the Fisher matrix before inverting, the resolution can be pushed slightly higher however the pixels appear noisier, as may be observed in the top right panel of Fig.~\ref{fig:MWay_maps}. The low SNR case proves, as in the simulation discussed above, that strong conditioning is required to recover the input distribution and that this leads to the exclusion of higher modes in the Fisher matrix. The level of the noise in this case is the same as in the one above. The monopole of the signal is very well recovered in both cases.

\section{Conclusions}\label{sec:concl}

We have presented a mapping algorithm for the LISA GW detector, which is treated as 3 correlated TDI channels. The method involves an iterative estimator which calculates the Fisher information matrix and gradient term by tracing over the channels, observation time, and frequencies efficiently. 
The tests discussed above clearly show there is a one-to-one relationship between the resolution $\ell_{\rm max}$ of the instrument and the frequency cutoff in the trace, and this is not particularly dependent on the input signal's spectral shape nor the spectral weighting assumed in the reconstruction. 
This frequency cutoff will naturally depend on both the nature of the signal and the level of the noise.
As may be read off of Fig.~\ref{fig:fmax}, the maximum angular resolution for LISA TDI 1.5 channels is $\ell_{\rm max}\sim 15$ in the best case scenario when the signal is loud and clear over the noise up to $f_{\rm max}=10^{-1}$ Hz. A possible candidate for this type of signal is a stochastic background of astrophysical origin from stellar mass black hole and neutron star binaries~\cite{Chen2018}.
The outlook is a lot less optimistic for lower frequency signals which are strong at $f < 10^{-2}$, for which $\ell_{\rm max} < 7$. An intermediate signal, which should peak around $f \sim 5 \times 10^{-2}$~\cite{Nelemans2001}, is the galactic background from binary white dwarfs. Considering this type of signal to be described by a simple power law in frequency, we have tested its reconstruction on the sky and have found that the strongest limit is set by the strong noise modes at high frequency. The reconstruction of low SNR signals is similarly limited, even if they are present at high frequencies, due to the shape and nature of the noise.\\
\indent In the tests presented above we assume that the size of the monopole sets both the SNR and the anisotropy level of the injected signal. However a more realistic modeling would see the introduction of two separate scales, a monopole $C_0$ and a typical anisotropy scale $C_1$. This would allow to test separately the detectability of background components with similar monopole values but different levels of anisotropy. For example, astrophysical extra-galactic backgrounds are expected to have a level of anisotropy $\delta\Omega/\bar{\Omega}=10^{-3}$ whereas cosmological backgrounds should present CMB-like ansotropies of $\delta\sim10^{-5}$~\cite{Contaldi:2016koz,Cusin:2018}. Hence, cosmological backgrounds from the early universe are expected to have a low SNR and a highly suppressed scale invariant anisotropy spectrum. These include inflationary backgrounds which seed primordial black holes \cite{Bartolo:2016ami}, highly non-Gaussian backgrounds~\cite{Ricciardone:2017kre,Dimastrogiovanni:2019bfl}, post-inflationary backgrounds due to strong first order phase transitions~\cite{Caprini2018} or cosmic defect networks~\cite{Auclair:2019wcv}; see \cite{Caprini:2019egz} for a full review. Map reconstruction for these background components will be quite challenging. The most promising candidate remains an astrophysical galactic background, hence we have carried out a preliminary study here. The LDC has produced mock time-domain data for a galactic background made up of thousands of superposed white dwarf binary inspirals which we plan to send through our pipeline soon.\\
 \indent The results obtained in this paper have been derived under the ideal assumption of a perfectly known noise PSD model. A natural expansion of this method would be to parametrise a given uncertainty in the noise model and performing parameter estimation on the noise and signal parameters simultaneously. This can happen by fitting noise parameters for the acceleration and interferometer components as seen in Equ.s~(\ref{noisecurv1}--\ref{noisecurv2}) for each cross-correlated channel or in each individual TDI channel. A technique for simultaneous reconstruction of the monopole component of the signal and a simplified mode of the noise power spectra, parametrised by acceleration and optical metrology system contributions only,  has been implemented in~\cite{adamscornish1,adamscornish2}, and more recently in~\cite{Caprini2019, Karnesis2019}. In addition, it has been shown that, under specific assumptions, a similar technique can also be efficient in the presence of idealised foregrounds~\cite{Pieroni:2020rob}. 

Another significant limit of this study is the assumption that the GWB intensity $I(f,\,\hat{\bm n})$ can be reduced to two independent components: a simple spectral shape and a constant pattern on the sky. We plan to explore the validity of this assumption in future work, studying specific GWB signals in the LISA band and analysing the time-domain generated LDC data containing anisotropic backgrounds. The results of this mapping algorithm may also be used to assess the detectability of different types of anisotropic backgrounds in the LISA band by studying the effective sky sensitivity of the detector and comparing it to the expected levels of anisotropy. This will be investigated further in~\cite{Angeloinprep}.\\

\section{Acknowledgements}
We highlight individual contributions to this manuscript. AIR: conceptualization, methodology, software, calculations, visualization, writing, project coordination. This was as a part of AIR's PhD thesis. MPi: conceptualization, methodology,
validation. CRC: conceptualization, methodology, supervision, funding. All others: analysis, validation, writing.
We thank Marc Lilley and Antoine Petiteau for useful conversations on the LISA response function and help setting up the pipeline in alignment with the LISACode format. We also thank Robert Caldwell for valuable comments.
AIR acknowledges support of an Imperial College Schr\"{o}dinger Fellowship. The work of CC and MPi was supported by Science and Technology Facilities Council consolidated grant ST/P000762/1. MPi was supported in part by the National Science Foundation under Grant No. NSF PHY-1748958. We would like to thank the IFT UAM-CSIC and the University of Padova for hosting the WG meetings where this project was born. MPi would like to thank the Kavli Institute for Theoretical Physics at UC Santa Barbara for the kind hospitality during part of this work. The work of G.C. has  received  funding  from  the European  Research  Council  (ERC)  under  the  European  Union's Horizon  2020  research  and  innovation  programme  (grant  agreement No 693024)  and from the Swiss National Science Foundation.  The work of GT is partially funded by STFC grant ST/P00055X/1.

\bibliographystyle{apsrev}
\bibliography{refs.bib}
\end{document}